\documentclass[12pt]{article}
\usepackage{epsfig}
\usepackage{array}
\usepackage{cite}
\usepackage{rotating}
\usepackage{amsmath}

\def\NPB{{\em Nucl. Phys.} B}
\def\PLB{{\em Phys. Lett.}  B}
\def\PRL{{\em Phys. Rev. Lett.}}
\def\PRD{{\em Phys. Rev.} D}
\def\ZPC{{\em Z. Phys.} C}

\def\IJA{{\em I.J.M.P.} A}

\def\EPJC{{\em Eur.  Phys. J C}}

\def\tb{\tan \beta}

\def\sab{\sin(\beta-\alpha)}
\def\s2ab{\sin^2(\beta-\alpha)}

\def\be{\begin{equation}}
\def\ee{\end{equation}}
\def\bea{\begin{eqnarray}}
\def\eea{\end{eqnarray}}

\def\gsim{\:\raisebox{-0.5ex}{$\stackrel{\textstyle>}{\sim}$}\:}

\def\bi{\bibitem}
\begin{document}
\begin{flushright}
IFT-2000-22\\[1.5ex]
{\large \bf hep-ph/0009201} \\

\end{flushright}

\begin{center}
{\bf \Large 
            THE LIGHT HIGGS WINDOW \\
            IN THE 2HDM AT GigaZ } 
\end{center}

\vskip 0.4cm
\centerline{MARIA KRAWCZYK
        }

\centerline{{\it Institute of Theoretical Physics, University of Warsaw, }}

\centerline{\it Warsaw, 00-681, Poland}
\vskip 0.5cm

\vskip 0.5cm
\centerline{PETER M\"{A}TTIG}

\centerline{\it Weizmann Institute, Rehovot, Israel}

\centerline{and}
\vskip 0.5cm
\centerline{JAN  \.ZOCHOWSKI}

\centerline{\it Faculty of Physics, Bia\l ystok University,}
\centerline{\it Bia\l ystok, Poland}

\begin{abstract}
The sensitivity to a light Higgs boson in the general 2HDM (II),
with a mass below 40 GeV, is estimated for an
future $e^+e^-$ linear collider operating 
with very high luminosity at the $Z$ peak 
(GigaZ). We consider a possible Higgs boson  production via 
the Bjorken process, the ($hA$) 
pair production,  
the Yukawa process $Z\rightarrow b{\bar b}h(A), \rightarrow \tau {\bar \tau}
h(A)$, 
and the decay $Z{\rightarrow}h(A)+\gamma$.
Although the discovery potential is considerably extended
compared to the current sensitivities, mainly from LEP, 
the existence of a 
$h$ or $A$ even with a mass of a 
few GeV cannot be excluded with 
two billion $Z$ decays.
The need to study the very light Higgs scenario 
at a linear $e^+e^-$ collider running at several hundred GeV
and the LHC is emphasised.

\end{abstract}


\section{Introduction}

Whereas for the minimal version of the 
Standard Model with just one scalar (Higgs) doublet,
data require the Higgs boson mass to be above
about 113 GeV~\cite{bib-OSAK2000},
no or less stringent limits can be set for more complicated sectors.
A fairly straight-forward extension of 
the Higgs sector in the Standard Model is, for example,
to assume two instead of one scalar doublets.
For a CP conserving model this implies
five physical bosons, two neutral scalars 
$h$ and $H$ (with $M_h<M_H$),
one neutral pseudoscalar $A$, and two charged Higgs bosons
$H^{\pm }$.
Apart from the masses of these bosons, the model is unambiguously
defined by specifying $\tan \beta $, given by
the ratio of the vacuum expectation values of the two doublets, 
the angle $\alpha $ describing the mixing in the
neutral scalars sector, and one 
Higgs boson self-coupling, say $g_{hH^+H^-}$ \cite{bib-hunter}.

In the context of the minimal supersymmetric
model (MSSM) a similar  Higgs sector 
with five physical Higgs bosons exists.
Relations between the parameters,
including the Higgs masses, 
are induced from the structure of the superpotential.
These relations reduce the number of independent parameters at tree level
to just two.
As a result, for some benchmark parameter sets of the MSSM,
e.g. maximal stop mixing scenario 
and $M_{SUSY}$ =  1 TeV, $m_{top}$ = 175 GeV,
$M_h$ and $M_A$ are constrained by data to be larger
than about 90 GeV and $\tan \beta $ to lie  
between about 0.5 and 2.3~\cite{bib-OSAK2000}.
The phenomenological consequences of the present measurements are quite 
different in the general 2HDM (Model II).
An analysis in this framework of the current constraints from LEP1 data, 
and from other experiments   can be found in \cite{bib-KZM,bib-chi2}.
In particular, present LEP data  are unable to rule out 
that either
$h$ is light but $M_A \gsim M_{Z}$ or vice versa.
Additional constraints for a very light $h$ or $A$,  
arise from  low energy experiments like $g-2$
of muons or searches for 
$\Upsilon$ decays into $M_h$ and $M_A$, see~\cite{g-2,keh}. 
Experiments at other than
LEP colliders have hardly any sensitivity to a light
2HDM Higgs boson~\cite{bib-MK-HERA}. 
Even after combining all existing experimental information 
there is still a large range 
in the parameter space to which no experimental sensitivity
exists and 
a window is left in the 2HDM(II) model for a
light Higgs boson extending down to even massless $h$ or  $A$.

Here we estimate the prospects for exploring the parameter space
of 2HDM (II) at the proposed TESLA~\footnote{
Although we will refer in this paper to the specific
TESLA scheme, our arguments apply equally well to
the other proposals for a Linear Collider~\cite{bib-NLC}.}
linear $e^+e^-$ 
collider~\cite{bib-TFS} running 
at the $Z$ peak, deemed GigaZ, with a luminosity of 
about two orders of magnitude
higher than LEP ~\cite{bib-GigaZ}.
In addition to the
higher luminosity improved charm and bottom
tagging capability will push the experimental sensitivity
to considerably 
smaller cross sections for the Higgs production processes.
We will study the potential for finding a light Higgs boson 
in particular for the process 
$Z{\rightarrow}h(A)+\gamma$, 
sensitive to both large and small $\tan \beta$ 
extrapolating our analysis of LEP data~\cite{bib-KZM},   
but also discuss   other important Higgs production processes at the $Z$.
At this stage no detailed simulation studies of future measurements 
are performed but the sensitivities
are estimated by extrapolating existing LEP measurements.
In addition we will briefly comment on the potential
signatures at the Linear Collider operating
at high energies of several hundred GeV and on implications
of the very light Higgs boson scenario at LHC.


\section{Some assumptions on experimentation at GigaZ}

We assume TESLA running 
at the $Z$ mass with an instantanous
luminosity of
7$\cdot $10$^{33}$ sec$^{-1}$cm$^{-2}$~\cite{bib-GigaZ} .
For a nominal year of 100 days this implies some
two billion $Z$'s to be produced, 
about a factor 500 more 
than what has been collected by each LEP experiment 
during five years of operation.
The forseen performance of a TESLA detector is basically
described in~\cite{bib-TFS}.

Pertinent for this study is
the potential for tagging bottom and charm quarks, 
which has been estimated in~\cite{bib-rhawkings},
and the photon energy resolution.
The possible efficiencies are given as a function of
the remaining background from the other flavours.
Without attempting to optimise the working point
of the tagging algorithm, we assume the following performance.
The bottom tagging
efficiency of 60$\% $ implies that only 2$\% $ of charm quarks
and 0.2$\% $ of light quarks will be retained.
Charm tagging has only been used with marginal efficiencies
and purities at LEP. 
However, the SLC experiment has shown the virtues
of a Linear Collider also for charm tagging. 
Experiments at TESLA are expected to tag charm with an
efficiency of 50$\% $ while accepting only 15$\% $ of bottom 
and 0.8$\% $ of light quarks.
No improvement is expected for tau identification.
Isolated photons should be easily identifiable and we
assume an energy resolution of 
$dE/E \ = \ 0.1/\sqrt {E}$~\cite{bib-TFS},
typically a factor 1.5-2 better than at LEP experiments.

We do not explicitely consider additional background sources.
However, beamstrahlung and underlying two-photon interactions
may become important in view of the very
small cross sections to be considered.


\section{Higgs production processes in $Z$ decays}

There are potentially four relevant mechanisms for the  lightest 
Higgs boson production at the Linear Collider running at the $Z$-peak.
They cover complementary regions in the 2HDM parameter space.
The Higgs strahlung $Z\rightarrow Zh$ has basically the same
experimental features as the Standard Model Higgs boson production.
In addition there is the pair production $Z\rightarrow Ah$
important for  $M_A \ + \ M_h \le \ M_Z$. 
Both of these are considered as the main production processes in the
framework of the MSSM.
The production yields of these 
processes are proportional to $ \s2ab $ and $\cos ^2 (\alpha - \beta )$,
respectively.

Both $A$ and $h$ can be singly produced in two other processes.
Since no relation between $M_h$ and $M_A$ exists, those  
processes are of particular interest within the 2HDM.
For $\tb >$ 1 
and low mass Higgs bosons, the Yukawa
processes $Z\rightarrow \tau ^+\tau ^- h(A),~b{\bar b}h(A)$, 
are  promising.
These processes depend on both $\sab $ and $\tb $ for 
$h$~\footnote{The corresponding coupling
is proportional to -$\sin \alpha / cos\beta= \sab-\tb \cos({\beta-\alpha}$)
for down type fermions. For up type fermions: $-\tb \rightarrow +1/\tb $.},
and for $A$ on just $\tb $ ($1/\tb $) for down (up) 
type fermions ~\cite{bib-Haber}. 
The radiative $Z$ decays $Z{\rightarrow}h(A)+\gamma$ 
is sensitive to both $\sab $ and values of $\tb $, 
see ~\cite{bib-KZM}.

Extrapolating from existing LEP studies
we will derive constraints on the parameters of the 2HDM for
masses between 5 and 40 GeV.
We will always refer to the
95\% confidence exclusion limit.
Note that the potential discovery of, say, a five standard
deviation Higgs signal excess over the Standard Model background
is only possible in a parameter space which is smaller than the
one for exclusions.


\subsection{Direct $Z$ decays into Higgs bosons}\label{sec:zzh}

The Higgsstrahlungs process
$$
Z\rightarrow Zh
$$ 
is the main channel to search
for a Standard Model Higgs boson at LEP.
In the 2HDM the yield is suppressed by $\sin ^2 (\alpha - \beta )$
compared to the Standard Model rate.
Current LEP data~\cite{bib-sine2LEP} yield minimum values of
$\sin ^2 (\alpha - \beta ) \sim {\cal O}$(0.006-0.01) for
5 $\le M_h \le $ 20 GeV and somewhat worse
$\sim {\cal O}$ (0.006-0.06) for 20 $< \ M_h \ < \ $40 GeV.
For the lowest masses  $M_h < $ 5 GeV
direct searches yield somewhat less restrictive limits of
about 0.01, however,
more restrictive limits of $\sim 5\cdot 10^{-3}$ are derived 
from the $Z$ line shape~\cite{bib-OPAL2hdm}.
The cleanest way to detect the Higgsstrahlungs process
for not too low Higgs masses is by tagging $Z$ decays with
electron or muon pairs,
each of which should yield at GigaZ 
some 6000$\cdot \s2ab $ events
of the type $e^+e^-h, \ \mu^+\mu^-h$ each for
$M_h \ \sim $ 5 GeV. 
The signature would then be two highly energetic leptons
with a mass close to $M_{Z}$ and some hadrons,
respectively two $\tau $ leptons.
Background is due to initial and final state (off shell) photon
radiation with the photon decaying into a pair of fermions.
In addition one has to consider the potential overlap
of an annihilation event with an event from two photon
interaction.
Other decays of the $Z$ can also be tagged with additional
experimental effort, in particular those 
into neutrinos and taus provide rather clean signals.
However, trigger efficiencies and backgrounds have to be
studied in more detail.
If hadronic decays of the $Z$ are included in the search,
jets from gluon radiation are the most important and rather
uncertain background. 

Compared to LEP the sensitivity
at the TESLA Linear Collider should improve by
$\sqrt {{\cal L}_{LC}/{\cal L}_{LEP}} \sim $ 20 
for a light Higgs boson and additional
factors if the Higgs boson decays into charm and bottom quarks. 
For $M_h \ge $ 10 GeV and $\tb >$ 1, the $h$ decay into
bottom is preferred.
For $M_h \ge $ 5 GeV and $\tb <$ 1,
$h$ decays mostly into
charm quarks.
Extrapolationg from current LEP limits we estimate that at TESLA
the range of 
$s^2_{LC} \ = \sin ^2 (\alpha - \beta ) > 5\cdot 10^{-4}$ 
could be covered for all masses below 
$M_h \sim $ 10 GeV.
For higher masses the limits should become less stringent,
reaching $\sim$0.005 for $M_h\sim $ 40 GeV.
This limit is almost independent of $\tb $.
In the following discussion we assume
$s^2_{LC}(M_h) =
 \sin ^2 (\alpha - \beta )_{LEP}(M_h)/20$,
where $\sin ^2 (\alpha - \beta )_{LEP}$ is the 
limit from LEP~\cite{bib-sine2LEP}.
If a light $h$ exists and is not found in the Higgsstrahlungs
process it implies that $h$ almost decouples from the $Z$.

If instead $M_h$ is large and the pseudoscalar $A$ is light,
the search for $Z\rightarrow Zh$ does not constrain
$\sin ^2 (\alpha - \beta )$.
Since no $ZZA$ coupling exists, direct $A$ production in
$Z$ decays can only occur via
$$
Z\rightarrow Ah
$$
whose yield is proportional to
$\cos ^2 (\alpha - \beta ) $.
If such a decay is kinematically possible, no
significant improvement over LEP limits can be expected from
GigaZ.
If $M_h \ > \ M_Z$ the mass of $A$ is unconstrained.


\subsection{The Yukawa Process}

The pseudoscalar $A$ 
can be singly produced by radiation
off heavy fermions.
The same is true for $h$ leading to complementary constraints
compared to the Higgsstrahlungs process.
Since the production yield is proportional to the
Yukawa coupling $f{\bar f}(h,A)$ we will refer to this
process as Yukawa process.
The experimentally most prominent decays are
$$
Z\rightarrow b{\bar b}h(A),\ \ \ \tau ^+\tau ^- h(A).
$$
For $\sin ^2 (\alpha - \beta ) $=0
the cross sections
scale with $\tan ^2 \beta$.
At LEP preliminary analyses have been presented for $A$ by 
ALEPH~\cite{bib-Ayuk} and for both $A$ and $h$ by 
DELPHI~\cite{bib-Dyuk}.

The dominant search channels are 
$$
\begin{array}{lcl}
Z\rightarrow & b{\bar b} ^-\tau ^+\tau, 
               \tau ^+\tau ^-\tau ^+\tau ^- & {\mathrm {for}}\  M_h\le  10\ 
                                              {\mathrm GeV} \\
Z\rightarrow & b{\bar b}b{\bar b} & {\mathrm {for} }\ M_h\ge  10\ 
                                              {\mathrm GeV}.
\end{array}
$$
The main Standard Model background will be $Z$ decays into
quarks, particularly bottom pairs with the additional emission 
of one or two hard gluons.
For very low Higgs masses also the background from overlapping
$Z$ decays and two - photon interactions may become
important.
Since the mass resolution of the $\tau ^+\tau ^-$ and
$b{\bar b}$ systems from a potential $h$ or $A$ decay is 
marginal, the signal has essentially to be derived from an
excess of the inclusive event yield of the candidate
topology.
At the Linear Collider with its higher luminosity and
higher bottom tagging efficiency such a search may finally
be limited by the knowledge of the irreducible background
processes.

For masses $M_{h,A} \ < \ $10 GeV and $\tb \ >$ 1,
the Higgs bosons will mainly decay into a pair of $\tau $
leptons.
Current analyses at LEP include hadronic $\tau $ decays.
To separate the Higgs signal from background
one has to understand yield and kinematic
features of low multiplicity gluon jets.
The uncertainty in their yield may be a limiting factor.
For masses larger than 10 GeV and $\tb \ >$ 1, the Higgs
boson will predominantly decay into a pair of bottom quarks.
In this case the main background will be due to $b{\bar b}g$
with the gluon splitting into $b{\bar b}$.
Its total cross section is $\sim $ 10pb, which has to be 
compared with a cross section of $\sim $ 0.01$\cdot \tan ^2\beta$ pb
for $M_{h,A}\ \sim \ $ 10 GeV, decreasing with the Higgs mass.
The background may be suppressed by selecting special
kinematical event properties.
However, one has to be aware that, apart from uncertainties
of 10-30$\% $ in the overall cross section for gluon splitting,
there are also uncertainties as to how gluons hadronise into
bottom particles.

An extrapolation to GigaZ is somewhat uncertain because of 
future theoretical developments and experimental
ideas to measure the yield and properties of
$g\rightarrow b{\bar b}$.
From the current understanding
we estimate that at GigaZ limits of 
$\tan \beta \ > \ {\cal O}(5)$ can be obtained for Higgs masses
between 5 and 10 GeV.
Above the bottom threshold it will be difficult to reach
sensitivities below $\tb $ = 10  for $M_{h,A} \ \sim \ 10 $ GeV,
respectively $\tb $ = 30 for $M_{h,A} \ \sim \ 40 $ GeV.
The corresponding limits are included in 
Figs.\ref{fig:sintan}-\ref{fig:expta}.

No LEP analysis exists on the Yukawa process for $\tb <$1.
In this range $h$ and $A$ would decay predominantly into charm quarks
since its coupling is proportional to 1/$\tb $.
Charm tagging at LEP is substantially less efficient than
bottom tagging.  
But even for the improved charm tagging at the Linear Collider
any sensitivity to such a decay will be extremely difficult
because of the significant irreducible background
$Z\rightarrow c{\bar c}c{\bar c}$.

\subsection{$Z{\rightarrow}h(A)+\gamma$}

This radiative decay of $Z$ proceeds for $h$ via loops of W - bosons 
and fermions.
In the Standard Model with one doublet the W - loop is by far the
dominant one, in the 2HDM it is suppressed by
$\s2ab $. The W - loop interfers negatively with the fermion loops 
in SM, in 2HDM with small $\sin (\beta -\alpha)$ destructive interference 
occurs only with up-type quarks (obviously with the largest effect for 
the top). 
The fermion  contributions   may be
enhanced  relative to the 
Standard Model by
$\sim \tan ^2 \beta $ (down type fermions) or
$\sim 1/\tan ^2 \beta $ (up type fermions, particularly top quark)
~\footnote{
Strictly speaking this proportionality holds only for small $\sab $.}.
For the radiative Z decay into the pseudoscalar $A$
only the fermion loops contribute.
In the 2HDM additional contributions are due to loops
of charged Higgs bosons for $h$ 
~\cite{bib-KZM}.
These are of minor importance for a heavy $M_H^{\pm}$ 
above 300 GeV as derived from the $b\rightarrow s\gamma$
yield~\cite{bib-bsg}, and a light Higgs boson $h$.

All LEP experiments have searched for radiative $Z$ decays
\cite{bib-ALEPH,bib-DELPHI,bib-L3,bib-OPAL}.
Their sensitivities are determined by the potential to observe a narrow
resonance over a smoothly varying, irreducible background
mostly due to photons emitted from the final state fermions
in standard $Z\rightarrow f{\bar f}$ decays.
Therefore, for all kinds of fermions, the luminosity and the
mass resolution of the $f{\bar f}$ system determine the
quality of the measurement.
The envisaged luminosity at GigaZ, being two orders of magnitude
higher than the one at LEP, will push the 
reach to this process significantly beyond existing limits.
An additional asset of the LC, depending on the decay mode of the Higgs
boson, is the increased capability of identifying fermion types.

The mass of the $f{\bar f}$ system, i.e. potentially the Higgs mass,
given by
$ M_{f{\bar f}} \ = \ E_{cm} \sqrt {1-2E_{\gamma}/E_{cm}}$,
is best determined,
at least for masses larger than about 30 GeV, through a precise measurement
of the photon energy $E_{\gamma}$.
For smaller masses the best resolution is obtained by a kinematic fit
to the three body system of fermions and the photon.
Here presumably no improved resolution will be possible compared to
current LEP measurements.
As discussed above,
because of the smaller beam pipe diameter and more sophisticated
micro vertex detectors,
the heavy quark tagging capability at the Linear Collider is largely
enhanced compared to LEP.
This will allow one to reduce the
background significantly if the Higgs decays into bottom or charm.

Assuming no signal will be observed, the limits $S_{LC}$ to be set at a
Linear Collider improve
 compared to those at LEP, $S_{LEP}$, according to
\begin{equation}
S_{LC} \ =\ S_{LEP} \sqrt {\frac {{\cal L}_{LEP}}{{{\cal L}_{LC}}}}
                    \sqrt {\frac {(dM_{f{\bar f}})_{LC}}
                                 {(dM_{f{\bar f}})_{LEP}}}
                    \frac {\epsilon_{\mathrm signal}^{LEP}}
                                 {\epsilon_{\mathrm signal}^{LC}}
                    \frac {\sqrt {\epsilon_{\mathrm bck}^{LC}}}
                                 {\sqrt{\epsilon_{\mathrm bck} ^{LEP}}}
\end{equation}
where ${\cal L}$ denotes the integrated luminosities at the
respective collider,
$dM_{f{\bar f}}$ the mass resolution of the ${f{\bar f}}$ system, and 
$\epsilon_{\mathrm signal}$ and 
$\epsilon_{\mathrm bck}$ the corresponding efficiencies for 
signal and background.
Thus, depending on the decay mode, one year of LC running will
boost the sensitivity
by factors of about 30 for $(q\bar {q})_{inclusive}$, i.e 
for all decays into quarks or gluons, 
and for decays into $\tau $ pairs, 36 for those into charmed quarks and
50 for those into $b\bar {b}$ pairs. 

\begin{figure}[p]
\center
\epsfig{file=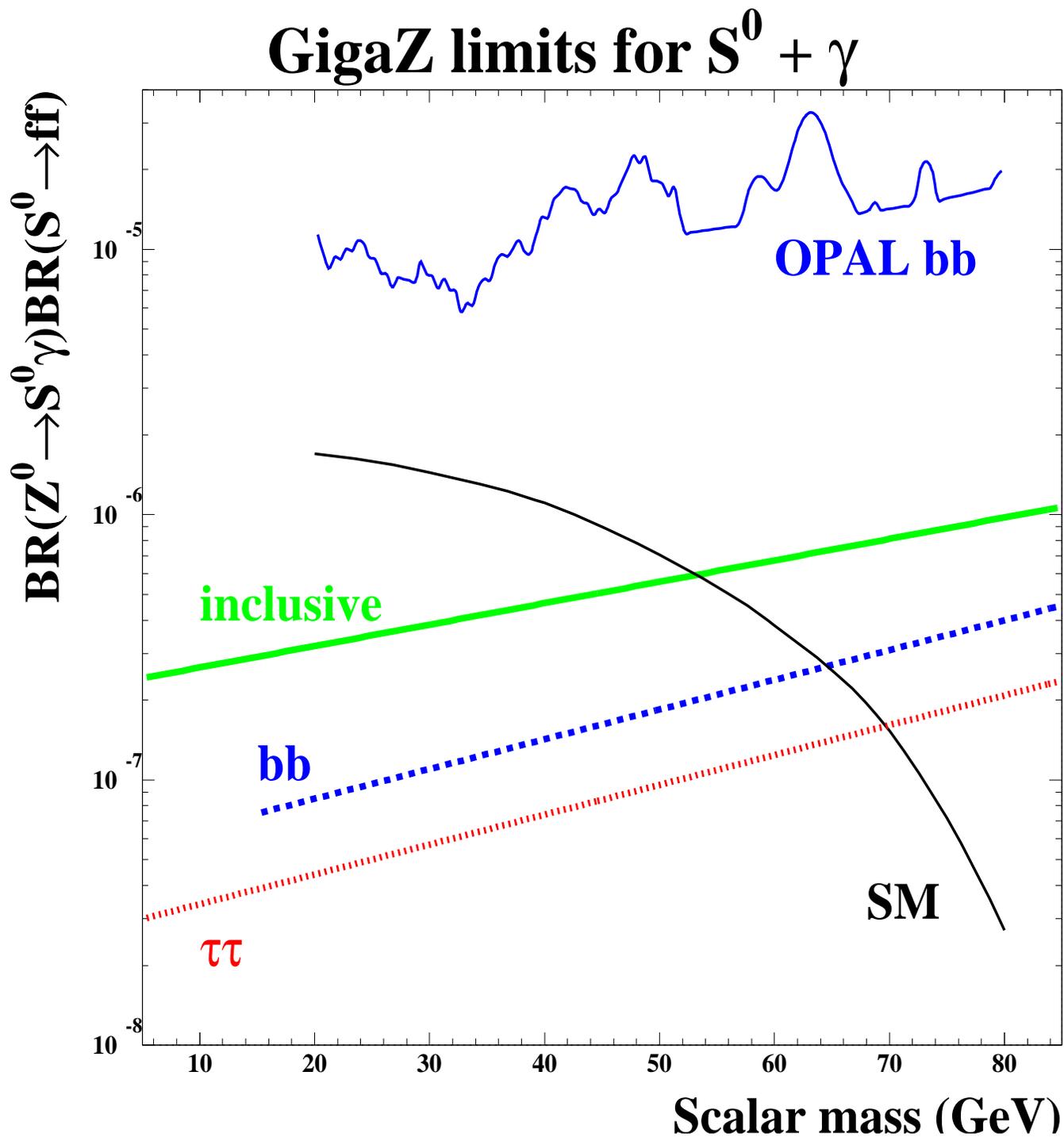,
        height=20cm,width=18cm}
\baselineskip 0.4cm
\caption{\sl Potential sensitivity on the branching ratio 
         $Z\rightarrow S+\gamma $ expected at GigaZ for various
         decay modes.
         The full (green) line shows the  
         limits for decays into any kind of quarks or gluons ({\sl inclusive}),
         the dashed (blue) into beauty quarks,
         and the dotted (red) into $\tau $ pairs.
         Also indicated are the current LEP reach for decays into
         bottom quarks~\cite{bib-OPAL} (narrow full/blue line)
         and the Standard Model prediction (black line).
         }
\label{fig:expt}
\end{figure}

The 95$\% $ confidence limits for the product branching ratio
${\mathrm {Br}}(Z\rightarrow (h,A) +\gamma)
{\mathrm {Br}}((h,A)\rightarrow {\bar f} f)$
can fairly well be described by a function of the form
\begin{equation}
S_{M_{h,A}} \ = \ K \cdot \exp [B \cdot (M_{h,A} - 30 GeV)],
\end{equation}
with $K$, $B$ constants depending on the Higgs decay mode and
$M_{h,A}$ in GeV.
The expected sensitivities are shown in Fig.~\ref{fig:expt}.
They will be used below to constrain the scalar $h$ and pseudoscalar $A$ 
production.
As a figure of merit also the rate  in the Standard Model
with just one Higgs boson is shown. 
In principle a Standard Model Higgs boson is in reach through
this process, but, of course,
for the accessible mass range it is already
excluded by LEP1 data.


\subsection{Constraints of the 2HDM parameters at GigaZ}

\begin{figure}[p]
\center
\epsfig{file=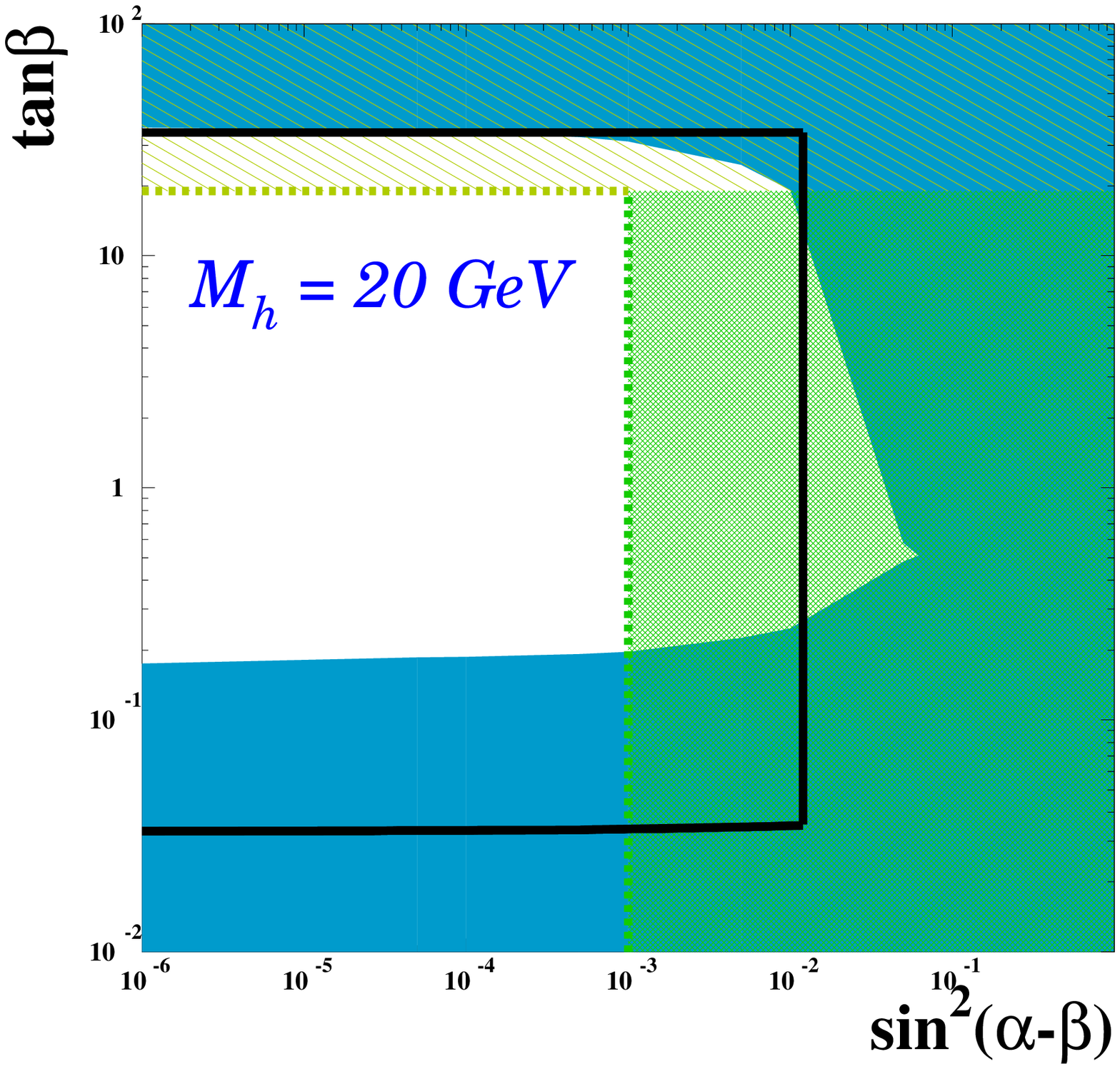,
        height=6.5cm,width=12cm}
\baselineskip 0.4cm
\caption{\sl Limits on  $\s2ab $ and $\tan \beta $
for a light scalar $h$ of mass $M_h$ = 6, 12, and 20 GeV at GigaZ.
The light hatched (yellow) region shows the exclusion from the Yukawa process,
the cross hatched (green) region the one from Higgsstrahlung,
the dark area depicts the exclusion range from $Z\rightarrow \gamma h$
(with mass of $H^{\pm}$=300 GeV).
Also shown are the current best limits from LEP (black line).
        }
\label{fig:sintan}
\end{figure}

\begin{figure}[p]
\center
\epsfig{file=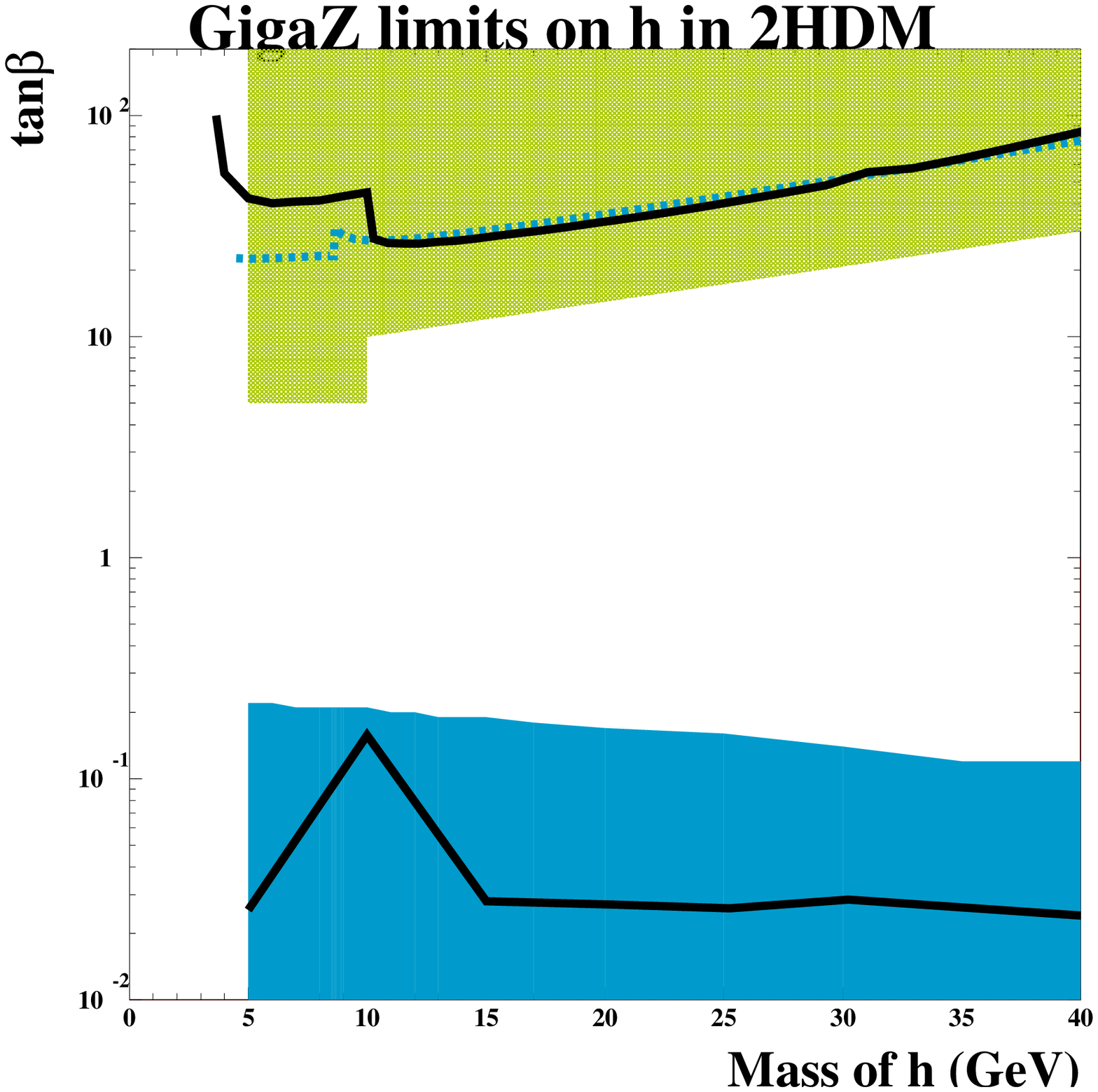,
        height=20cm,width=18cm}
\baselineskip 0.4cm
\caption{\sl Potential exclusion range at GigaZ for the light scalar $h$ 
         as a function of its mass and $\tan \beta$ (for $\sab$=0).
         The upper light (green) shaded region is excluded by the 
         Yukawa process,
         the lower dark (blue) shaded region due to $Z\rightarrow h\gamma $.
         The upper limit from $Z\rightarrow h\gamma $ for $\tb >$1 is
         indicated by the (blue) dotted line.
         Also shown is the current best limits obtained at LEP 
         (black lines - upper from the Yukawa process
          and lower from the radiative Z decay).  
        }
\label{fig:expth}
\end{figure}

\begin{figure}[p]
\center
\epsfig{file=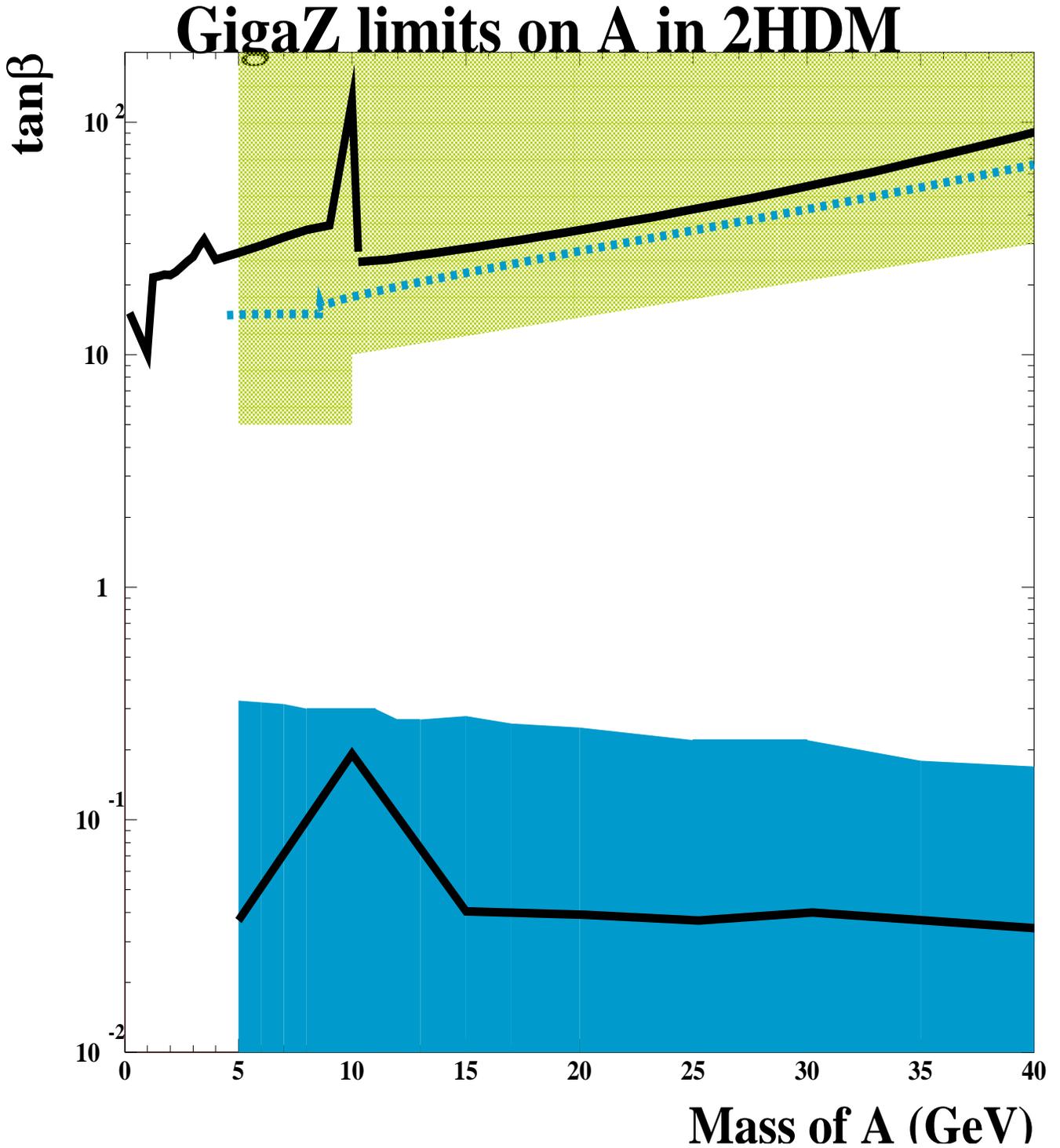,
        height=20cm,width=18cm}
\baselineskip 0.4cm
\caption{\sl Potential exclusion range at GigaZ for the light scalar $A$ 
         as a function of its mass and $\tan \beta$.
         The upper light (green) shaded region is excluded by the 
         Yukawa process,
         the lower dark (blue) shaded region due to from $Z\rightarrow 
         A\gamma $.
         The upper limit from $Z\rightarrow A\gamma $ for $\tb >$1 is
         indicated by the (blue) dotted line.
         Also shown is the current best limits obtained at LEP  (black lines-
         upper from the Yukawa proces 
         and lower from the radiative Z decay).
        } 
\label{fig:expta}
\end{figure}

The expected sensitivity ranges at GigaZ for $h$, $A$ are 
depicted
in Figs.~\ref{fig:sintan}-\ref{fig:expta}.

In Fig.~\ref{fig:sintan} the potential exclusion range for a
scalar Higgs of $M_h$ = 6, 12, or 20 GeV is shown in the 
($\tb $, $\s2ab $) plane.
These figures exhibit the complementarity of the considered  processes.
The Higgsstrahlung $Z\rightarrow Zh$ bounds the value of
$\s2ab $, excluding the region indicated at the right of the 
figures.
For the remaining allowed region of very small $\s2ab $,
the Yukawa process sets a lower limit on $\tb $ excluding
the upper part of the figures.
The radiative process $Z\rightarrow h +\gamma $ is sensitive
to both $\tb $ and $\s2ab $.
Its exclusion range is unique for $\tb < $1 and 
$\s2ab \ \rightarrow $ 0.
It should be noted that if a light $h$ is found at GigaZ
the complementarity of these processes
allows one to significantly constrain the parameters in the
2HDM.
Also shown in Fig.~\ref{fig:sintan} are the current constraints 
from LEP, as discussed in previous sections. Obviously the 
GigaZ can significantly extend the excluded region of the parameter space.

The potential exclusion region for $h$ in the 
($\tb $, $M_h $) plane is shown in Fig.~\ref{fig:expth}.
Here constraints from $Z\rightarrow h + \gamma $ were obtained
under the most conservative
assumptions of $\sab$=0 and a mass of the charged Higgs boson 
of 300 GeV. 
In the range $\tb >$1 the most constraining result comes
from the Yukawa process,
however does not reach below $\tb \sim $ 5 for $M_h<$ 10 GeV 
and $\tb $ = 10 - 30 for higher masses.
For $\tb <$ 1 limits are due to the radiative Z decay and are
of ${\cal O}(0.2-0.1)$, rather independent of $M_h$.
The sensitivity of the $Z\rightarrow h + \gamma $ decay
for $\tb >$1 is indicated by the dashed line and is seen
to be less constraining than the Yukawa process.
Also shown is the current limit from LEP revealing
a substantial gain at GigaZ.
However, if the $h$ decouples from the $Z$, i.e. 
$\s2ab \sim$0 a sizeable parameter space for the scalar Higgs $h$
remains uncovered - 
a light neutral Higgs boson in the 2HDM remains a possibility.

A similar picture emerges for the pseudoscalar $A$
shown in Fig.~\ref{fig:expta}.
The main difference is that its sensitivity range is 
independent of $\s2ab $.
The Yukawa process yields similar  exclusion potential for
$h$ and $A$ in the region $\tb >$1.
Also in this case the radiative decay has a smaller reach 
than the Yukawa process.
The sensitivity to $A$ is larger than the one for $h$
because of the absence of the negative interference of 
 top and $W$ loops.
Out of the same reason also the exclusion in the $\tb <$1
region is somewhat better ${\cal O}(0.3-0.2)$.
Combining all constraints, also a light $A$ cannot be
excluded for all 2HDM parameters at GigaZ.
As in the case for $h$, GigaZ improves significantly over
LEP, whose limits are indicated by the full line.


\section{The High Energy frontier}

Up to now we have only considered the potential 
of finding a light Higgs boson $h$ or $A$
at the $Z$ peak.
Although the sensitivity to a light
$h$ or $A$ can be significantly extended at GigaZ, 
even two billion produced $Z$'s are not enough to
definitely find or rule out a very light 
$h$ or $A$ in the framework
of the 2HDM (II). 
On the other hand, 
if such a light Higgs boson exists 
  but is not found, the sensitivity reach at GigaZ would imply 
${\cal O}(0.2) \le \tan \beta \ \le {\cal O}(20)$.
If the light Higgs boson is a scalar $h$, then in addition
$\sin ^2 (\alpha - \beta ) \ \sim \ 0$, i.e.
the $h$ decouples from the $Z$.

Apart from the GigaZ option other future projects are under
consideration.
Potentials for the discovery of a light Higgs boson in 2HDM (II)
in $\gamma \gamma $, and $\mu ^+\mu ^-$ colliders
are discussed in \cite{bib-futcoll}.
Improved constraints may also follow from 
low energy high precision experiments, e.g. the E821 experiment for 
$g-2$ for muon at BNL \cite{bib-new-g2}.
All those experiments can increase the sensitivity, but
they cannot cover the whole parameter space of the 2HDM
for a light Higgs boson.
Thus, if one does not observe the Higgs boson, one
will not be able to rule out the light Higgs scenario.  
Here we want to briefly comment on possibilities
for exploring the 2HDM with the higher energies available
both at TESLA  and the LHC.
In most cases a very light Higgs scenario, discussed in this paper,
has been neglected in prospective studies
for these machines.
More detailed considerations are needed for firm conclusions
on the potential for probing the very light Higgs scenario. 
We will suggest some lines of studies in the following
paragraphs.

At a high energy $e^+e^-$ collider the
direct production of a light $h$ or $A$ 
via the pair production $hA$ 
with either $A$ or $h$ being light and the other Higgs
boson from the pair having a mass of a few hundred GeV, 
and the 
$t{\bar t}(h,A)$ Yukawa process may probe new regions
of the 2HDM parameter space. 
In the first reaction
the final decay products can be, for example, $W^+W^-(A,h)$, 
$ZZ(A,h)$. 
The second process exploits the strong Yukawa coupling
to top quarks.
The emission of a $h$ or $A$ from top quarks can be searched
for with the Higgs boson decaying into pairs of bottom, charm or tau.
Here again one has to discriminate against background from
gluon splitting into quarks.
A related signal is a possible
enhancement at the $t{\bar t}$ threshold 
due to virtual Higgs boson exchange~\cite{bib-tthresh}.
To understand the precision and reach of such processes,
more detailed studies including experimental
effects and updated parameters have to be 
performed~\footnote{
Within a CP non conserving 2HDM the sensitivity at a 
500 - 800 GeV Linear Collider has been theoretically
considered in~\cite{bib-CPviol}. 
Also their study, 
not particularly focused on the very light Higgs scenario,
indicates that from the Bjorken process, 
pair production and the Yukawa process the whole parameter
space cannot be covered.}.

There are plenty of channels at
$pp$ and $p\bar p$ colliders in which a light Higgs boson may be produced with
a sizeable event yield.
However, it is not obvious if those can be isolated from
the formidable background.
Most studies have been performed for 
$M_{h,A}>$100 GeV and indicate that it is
increasingly difficult to detect a
Higgs signal the smaller the Higgs mass is.
On the other hand detailed studies for masses 
of a few GeV are missing.
The only analysis so far at $pp$ colliders with relevant 
limits has recently performed by the
CDF collaboration and yields limits 
within the MSSM from the Yukawa process 
$b{\bar b}h$ for large $\tb $ and 
for Higgs masses larger than 70 GeV~\cite{bib-CDFhiggs}.

As discussed in~\cite{bib-chi2} a global fit to current precision EW 
measurements performed in the 2HDM(II)
allows the existence of a very light $h$ or $A$, with mass even below 20 GeV,
while the other Higgs bosons may have masses of several hundred GeV.
The detection of such massive Higgs bosons may be possible 
at high energy colliders as TESLA or the LHC.
These heavy Higgs bosons may reveal the first signal of the 2HDM sector
and open up the window for the exploration of a very light Higgs boson,
discussed in this paper.

Analyses in the framework of the MSSM have shown a good
sensitivity both at TESLA and LHC
to such massive bosons if they decay directly
into gauge bosons or $b{\bar b}$ pairs~\cite{bib-TESLAmssm,bib-LHCmssm}.
These studies also basically apply to the 2HDM.
The existence of a light Higgs boson in the general 2HDM, 
however, opens up the
possibility of cascade decays which result into more 
involved decay patterns.
Decays like $H \rightarrow h h $, $A\rightarrow Zh $,
$H^{\pm }\rightarrow W^{\pm }h $ (for a light $h $),
respectively
$h \rightarrow AA$, $h \rightarrow ZA $ or $H^{\pm }\rightarrow W^{\pm }A $
(for a light $A$) are more complicated to reconstruct.
The feasibility of identifying these decays requires
additional studies.
Those performed in the LHC framework,
again at this stage only in the MSSM, indicate that it may
be difficult to cover the whole parameter space.
At TESLA these events could have quite distinct features and be
relatively simply to identify.

In conclusion we want to reemphasise that the case of a very
light Higgs boson, i.e. with a mass of 40 GeV or below - even  
only a few GeV, is not closed.
A $Z$ factory like GigaZ producing two billion $Z$'s 
per year will
allow one test a significantly larger
parameter space of the 2HDM compared to what has yet
been probed.
However, it seems unlikely that the whole space can be covered.
A window for a very light Higgs boson remains.
It may be possible to study the whole range with the high energies
provided at TESLA and LHC, however, most studies so far
have not considered very light Higgs bosons.
More definite conclusions can only be reached if detailed
analyses are performed taking into account the light mass
Higgs window. 

\section{Acknowledgements}

We thank Piotr Chankowski for important and enlightening 
discussions on parameters of 2HDM. 
 MK thanks   Michael Kobel 
 for informing of and sending the paper \cite{keh}.
We are grateful to J. K\"uhn and M. Je\.zabek 
to make us aware of the 
sensitivity of the $t\bar t$ threshold to the light Higgs.
MK would like to thank Padova University for warm hospitality 
during the preparation of this paper.
This work was partly supported by Polish Committee for 
Scentific research, grant No 2P03B01414. 


\newpage

\begin{figure}
\center
\epsfig{file=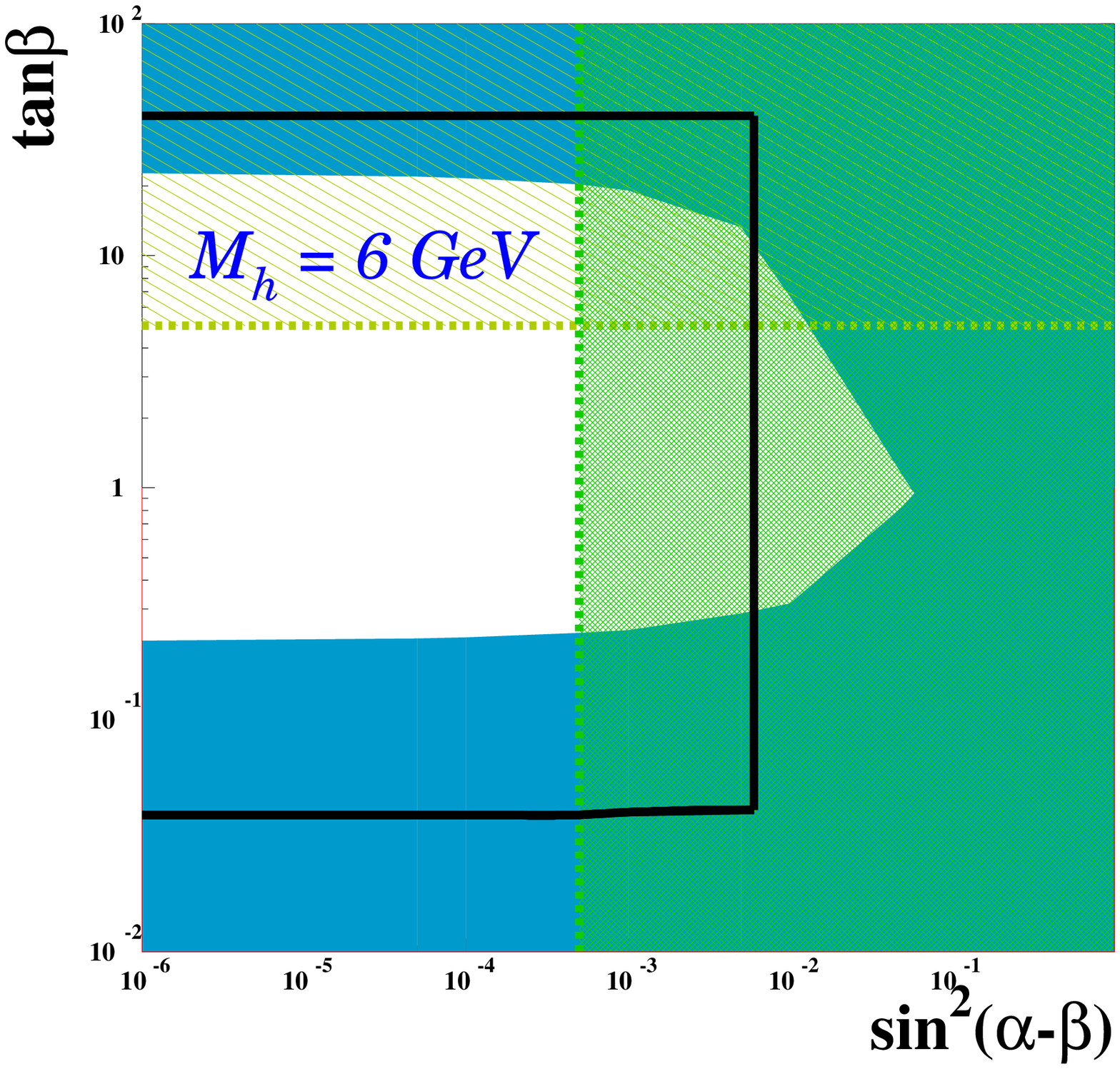,
        height=6.5cm,width=12cm}

\label{fig:m6}
\end{figure}

\vspace{-1.6cm}

\begin{figure}
\center
\epsfig{file=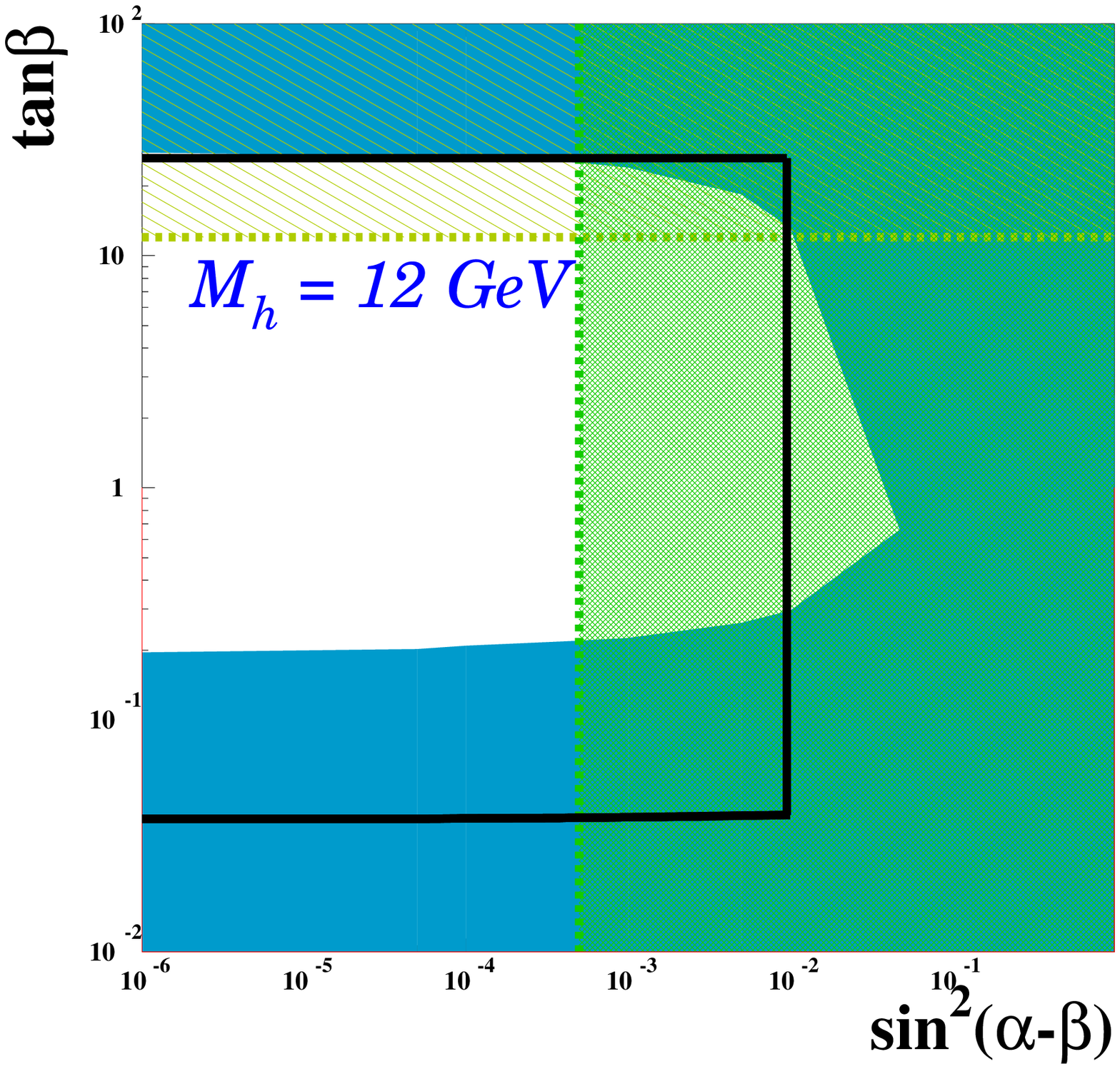,
        height=6.5cm,width=12cm}
\label{fig:m12}
\end{figure}

\vspace{-1.6cm}

\newpage

\newpage

\newpage

\end{document}